\documentclass[preprint,12pt]{elsarticle}

\usepackage{graphicx}

\usepackage{amssymb}

\begin{document}

\begin{frontmatter}

\title{Homotopy analysis method for stochastic differential equations}


\author[l1,l2]{Maciej Janowicz}
\author[l2]{Filip Krzy\.zewski}
\author[l1]{Joanna Kaleta}
\author[l1]{Marian Rusek}
\author[l1,l2]{Arkadiusz Or{\l}owski}

\address[l1]{Faculty of Applied Informatics and Mathematics, 
Warsaw University of Life Sciences - SGGW,
ul. Nowoursynowska 159, 02-786 Warsaw, Poland}
\address[l2]{Institute of Physics of the Polish Academy of Sciences,
Aleja Lotnik\'ow 32/46, 02-668 Warsaw, Poland}

\begin{abstract}
The homotopy analysis method known from its successful applications
to obtain quasi-analytical approximations of solutions of ordinary
and partial differential equations is applied to stochastic differential
equations with Gaussian stochastic forces and to the Fokker-Planck 
equations. Only the simplest non-trivial examples of such equations
are considered, but such that they can almost immediately be translated
to those which appear in the stochastic quantization of a nonlinear
scalar field theory. It has been found that the homotopy analysis
method yields excellent agreement with exact results (when the latter 
are available) and appears to be a very promising approach in the 
calculations related to quantum field theory and quantum statistical
mechanics.

\end{abstract}


\end{frontmatter}


\section{Introduction}
\label{Intro}

The unprecedented recent growth of computing power of modern machines as well as the development
of the software has made it possible to substantially enhance the quantitative predictive
capabilities of physical research. Nevertheless, it appears that the role played by analytical
techniques to solve mathematical problems of science and engineering has by no means diminished.
And the development of symbolic algebra systems
has 
enabled the analytical methods to become more reliable and efficient.
Among those analytical methods one can mention a particularly efficient approach recently developed 
by Liao \cite{Liao1,Liao2,Liao3,LT,Liao4,Liao5} called by that author the ``homotopy analysis method".
One of the most significant features of the homotopy analysis method is its independence of the presence

of any small parameter. The approximate solution has the form of a series expansion.
Usually, it is required that the series is truncated at a rather high order. However, skillful
choice --- e.g., on using variational tools --- of some additional control parameters 

(the presence of which the method admits) allows one to obtain meaningful results even in low orders, 
as reported by Marinca and coworkers \cite{MHN,MH1,MH2}.

The purpose of this paper is to demonstrate usefulness of the homotopy analysis method in the case
of stochastic differential equations. To illustrate the quality of approximation provided
by Liao's method, we use a single and very simple stochastic differential equation, all statistical
properties of its solution being easily obtained from the corresponding Fokker-Planck (Kolmogorov 
forward) equation. Our choice of working example has been, however, also motivated by the far-reaching
analogy with the finite-temperature quantum field theory in the stochastic quantization framework.

The main body of this work is organized as follows. Section 2 contains a short
description of the Liao as well as Marinca approaches to obtain series solutions of the 
differential equations. In Section 3 such a series is obtained for a stochastic differential
equation related to the so-called ``zero-dimensional'' quantum field theory.
Section 4 is devoted to the solution of the corresponding Fokker-Planck equation
while Section 5 contains several final remarks.

\section{Homotopy analysis method by Liao}
\label{Dwa}

Let us consider an arbitrary system of differential equations (linear or non-linear, 
ordinary or partial, homogeneous or not) of the form

\begin{equation}
\label{n1}
{\mathcal N}_{a}(u, t) = 0
\end{equation}

\noindent where the index $a$ enumerates the equations of the system, $u$ is the 
vector of dependent variables, and $t$ represents the set of independent variables. 
The family of solutions will of course not change if we multiply the left-hand side 
by a parameter $q$, $q \neq 0$.
Alongside (\ref{n1}) let us also consider a one-parameter family of systems of the form:

\begin{equation}
\label{n2}
[(1 - q) {\mathcal L} (u(t) - u^{(0)}(t))]_{a} + q {\mathcal N}_{a}(u, t) = 0, 
\end{equation}

\noindent where ${\mathcal L}$ is a linear operator such that ${\mathcal L} u = 0$ 
can easily be solved while $u^{(0)}$ is an initial guess of the solution. 
Very often $u^{(0)}$ is chosen either to satisfy ${\mathcal L} u^{(0)} = 0$ or simply $u^{(0)} = 0$. 
Thus, for $q = 0$ we have to do with an easy to solve linear system while for $q = 1$ 
we obtain the initial system.
The idea of Liao has been to solve the system (\ref{n2}) by expanding $u$ into a power
series with respect to $q$ 

$$
u = u^{(0)} + q u^{(1)} + q^{2} u^{(2)} + ...
$$

and set $q = 1$ at the end of calculations. Obviously, 
if that were all, we would merely obtain a variant of perturbation expansion with artificial 
``small'' 
parameter introduced by hand, even though ${\mathcal L}$ need not be contained 
in ${\mathcal N}$ and $u^{(0)}$ need not be a solution to ${\mathcal L} u = 0$. 
However, Liao observed that $q {\mathcal N}_{a}(u, t)$ can still be multiplied by a parameter 
$\xi$ and function $h(q)$, the latter being analytical in $q$ without changing the solution 
of (\ref{n2}) for $q = 1$. This gives the system:

\begin{equation}
\label{n3}
[(1 - q) {\mathcal L} (u(t) - u^{(0)}(t))]_{a} + \xi q h(q) {\mathcal N}_{a}(u, t) = 0, 
\end{equation}

The function $h(q)$ can be chosen in such a way as to, e.g.,
redefine a ``real" expansion parameter (which is not explicitly visible in (\ref{n1}) to make it smaller, 
while $\xi$ can be chosen at the end of calculations to improve the overall convergence.
In fact the proper choice of the parameter $\xi$ can and usually does improve the quality
of approximate solution quite dramatically, and its apparently insignificant presence
facilitated the spectacular success of the method.
Further considerable progress has been achieved by Marinca and Herisanu who proposed to optimize the choice 
of the coefficients in the expansion

$$
h(q) = h_{0} + h_{1} q + h_{2} q^2 + ... 
$$

by minimization of residual error.
Even with those amendments, the method, in generic case, suffers from the presence 
of secular terms if appplied to resonant systems so that the expansion is not uniformly valid. 
However, any known mechanism to eliminate the secular terms can be employed to augment the power of
Liao's approach, e.g. the Lindstedt-Poincare method. Here, we will use a variant
of the Lindstedt-Poincare method by observing that the operator ${\mathcal L}$
can itself be an analytical function of $q$. This will allow us to eliminate (at least
in the sense of expectation values, please see below) the terms which would correspond
to the seculat terms in ordinary differential equations.

\section{Application to stochastic differential equation with third-order nonlinearity}

We consider the following non-linear stochastic ordinary differential equation:

\begin{equation}
\label{row1}
\frac{d \phi(s)}{d s} = -\alpha \phi(s) - \frac{1}{6} \lambda \phi(s)^{3} + f(s),
\end{equation}

where the stochastic ``force" $f(s)$ is a Gaussian and Markovian and its statistical
properties are given by:

$$
\langle f(s) \rangle = 0,
$$

$$
\langle f(s) f(s^{\prime}) \rangle = 2 v \delta (s - s^{\prime}),
$$

where $v$ is a real parameter larger than zero and the sharp brackets denote 
expectation values.

Physically, the above equation describes heavily overdamped anharmonic oscillations 
of a Brownian particle. It has been used by Bender and coworkers to introduce
strong-coupling expansion in classical statistical mechanics \cite{BCGRS}.
However, we would like to emphasize here another motivation: Eq.(\ref{row1}) can
be obtained as:

\begin{equation}
\label{row2}
\frac{d \phi(s)}{d s} = -\frac{d S_{E}}{d \phi} + f(s),
\end{equation}

where 

\begin{equation}
\label{row3}
S_{E} = \frac{1}{2} \phi a \phi + \frac{1}{4!} \phi^{4} 
\end{equation}

is sometimes called the action of a ``zero-dimensional'' scalar meson field theory
with quartic self-interaction. This is because the full four-dimensional
action (in imaginary or Euclidean time $\tau = -i t$) has the form:

\begin{equation}
\label{row4}
S_{E} = \int d \tau d^{3} r \left( \frac{1}{2} \Phi({\bf r}, \tau) L_{KG} \Phi({\bf r}, \tau) + 
\frac{1}{4!} \Phi({\bf r}, t)^{4} \right), 
\end{equation}

where $L_{KG}$ is the linear Klein-Gordon operator written in the imaginary time.

The fact that all statistical properties of the quantized meson field thery can be obtained
from stationary ($s \rightarrow \infty$) solutions of the system:

\begin{equation}
\label{row5}
\frac{d \Phi(s)}{d s} = -\frac{\delta S_{E}}{\delta \Phi} + f({\bf r}, \tau, s),
\end{equation}

where $\delta/\delta \Phi$ is the variational derivative, has been established 
in \cite{PW,DH,Namiki,HR,Gozzi}. The transition from (\ref{row4}) to (\ref{row5})
is usually called ``stochastic quantization''. 

By stationary solution of (\ref{row1}) or ({\ref{row5}) we mean the solution
obtained for large $s$ so that all the initial correlation have died out
and it is sufficient to take into account only the special solution
of the inhomogeneous equation.
 
The fact that (\ref{row3}) is the zero-dimensional caricature of (\ref{row4})
($S$ becomes merely a simple function instead being a functional of the field $\Phi$)
is evident, and the same is true about ({\ref{row1}) and (\ref{row5}).
The way any special method is applied to solve (\ref{row5}) is a direct
generalization of its application to (\ref{row1}).

Let us now observe that Eq. (\ref{row1}) admits an exact solution in the sense
that all moments of $\phi$ can be easily obtained. Indeed, the corresponding
Fokker-Planck equation for the distribution function $\rho(\phi, s)$ takes the form:

\begin{equation}
\label{row5a}
\frac{\partial \rho(\phi, s)}{\partial s} = v \frac{\partial^{2} \rho(\phi, s)}{\partial \phi^{2}} + 
\frac{\partial}{\partial \phi} \left[ \left( a \phi + \frac{1}{6} \lambda \phi^{3} \right) 
\rho(\phi, s) \right],
\end{equation}

which admits the stationary ($s$-independent) solution:

\begin{equation}
\label{row5b}
\rho = \rho(\phi) = N \exp \left[ -\frac{1}{v} \left( \frac{1}{2} \alpha \phi^{2} + 
\frac{1}{4!} \lambda \phi^4 \right) \right],
\end{equation}

where $N$ is a normalization constant obtained from the normalization condition:

$$
\int_{-\infty}^{\infty} \rho(\phi) d \phi = 1,
$$

so that

$$
N = \left[ \sqrt{3} \exp\left( \frac{3 \alpha^{2}}{4 \lambda v} \right) \sqrt{\frac{\alpha}{\lambda}} 
K_{1/4}\left( \frac{3 \alpha^{2}}{4 \lambda v} \right) \right]^{-1}
$$

This way we obtain, in particular:

$$
\langle \phi^{2} \rangle_{st} = 
\frac{e^{-\frac{3 \alpha ^2}{4 \lambda v}} \left(16 \sqrt{3}
   \sqrt{\lambda v} \Gamma \left(\frac{7}{4}\right) {}_ 1 F_{1}
   \left(\frac{3}{4};\frac{1}{2};\frac{3 \alpha ^2}{2 \lambda v
   }\right)-9 \sqrt{2} \alpha  \Gamma \left(\frac{1}{4}\right) {}_1 F_{1}
   \left(\frac{5}{4};\frac{3}{2};\frac{3 \alpha ^2}{2 \lambda v
   }\right)\right)}{3 2^{3/4} \sqrt[4]{3} \sqrt{\alpha } \lambda
   ^{3/4} K_{\frac{1}{4}}\left(\frac{3 \alpha ^2}{4 \lambda v
   }\right)},
$$



where the braces $\langle ... \rangle st$ denote stationary expecation value,
${}_1 F_{1}$ is the hypergeometric function, $\Gamma$ is the Euler Gamma function,
and $K_{1/4}$ is the modified Bessel function of the second kind.

Below, we pretend that we do not know the above exact expressions. In order to test
whether the homotopy analysis method is suitable for the stochastic
differential equations, we shall attempt to solve (\ref{row1}) approximately.

Following Liao, alongside (\ref{row1}) we consider the equation: 

\begin{equation}
\label{row6}
(1 - q) (\frac{d}{d s} + \nu(q)) \phi(q, s) + q c(q) 
\left[ \frac{d \phi(q, s)}{d s} + \alpha \phi(q, s) + \frac{1}{6} \lambda \phi(q, s)^{3} \right] = 0,  
\end{equation}

in which we expand $\phi$, $\nu$, $c$ as:

$$
\phi(q, s) = \sum_{n \geq 0} \phi_{n} q^{n},
$$

$$
\nu(q) = \sum_{n \geq 0} \nu_{n} q^{n},
$$

$$
c(q) = \sum_{n \geq 0} c_{n} q^{n}.
$$

However, in order not to overburden the expansions, we will equate all $c_{n}$ with non-zero $n$ to zero.
In the end of calculation, we shall set $q$ = 1 and 

\begin{equation}
\label{row7}
\sum_{n \geq 0} \nu_{n} q^{n} |_{q = 1} = \alpha.
\end{equation}

The last condition means that for $\lambda = 0$, $q = 0$, we obtain the ``unperturbed", linear problem

$$
\frac{d \phi}{d s} = -\alpha \phi + f.
$$

Let us notice here that the choice of the constraints in Eq. (\ref{row7}) is only
the simplest of many possibilities. In specific cases, other choices may be preferrable.
As mentioned before, there is no obvious way to get $\nu_{1}$, $\nu_{2}$, etc. In the case
of deterministic resonance problems we would choose $\nu_{n}$, $n \geq 1$ in such a way
as to avoid the secular terms. Here, no secular terms appear. Below we propose
a scheme to choose $\nu_{n}$ which seems to us simple and intuitive. Another
choice is discussed in Section 4.

In the zeroth-order we obtain, of course:

\begin{equation}
\label{row8}
\left( \frac{d}{d s} + \nu_{0} \right) \phi_{0}(s) = f(s),
\end{equation}

so that 

\begin{equation}
\label{row9}
\phi_{0}(s) = \phi_{0}(0) e^{-\nu_{0} s} + 
\int_{0}^{s} e^{-\nu_{0} (s-s^{\prime})} f(s^{\prime}) d s^{\prime} 
\end{equation}

For large $s$ we have 

\begin{equation}
\label{row10}
\phi_{0}(s) = \int_{0}^{s} e^{-\nu_{0} (s-s^{\prime})} f(s^{\prime}) d s^{\prime} 
\end{equation}

so that, for large $s, s^{\prime}$ the correlation function 
$\langle \phi_{0}(s) \phi_{0}(s^{\prime})$ is given by:

\begin{equation}
\label{row11}
\langle \phi_{0}(s) \phi_{0}(s^{\prime}) \rangle = 
\frac{v}{\nu_{0}} \left( \exp(-\nu_{0} |s - s^{\prime}| ) - \exp(-\nu_{0} (s + s^{\prime}) \right),
\end{equation}

which for $s = s^{\prime}$ becomes:

\begin{equation}
\label{row12}
\langle \phi_{0}(s)^{2} \rangle = \frac{v}{\nu_{0}} \left( 1 - \exp(-2 \nu_{0} s) \right), 
\end{equation}

and, naturally, $\langle \phi^{2} \rangle_{st} = v/\nu_{0}$.

In the first order we need to solve: 

\begin{equation}
\label{row13}
\left(\frac{d}{d s} + \nu_{0} \right) \phi_{1}(s) + (\nu_{1} + c_{0} (a - \nu_{0})) \phi_{0}
+ \frac{1}{6} c_{0} \lambda \phi_{0}(s)^{3} = 0.
\end{equation}

The solution for large $s$ is given by:

\begin{equation}
\label{row14}
\phi_{1}(s) = -\int_{0}^{s} e^{-\nu_{0}(s-s^{\prime})} \left[ 
(\nu_{1} + c_{0} (\alpha - \nu_{0}) ) \phi_{0}(s^{\prime}) + 
\frac{1}{6} c_{0} \lambda \phi_{0}(s^{\prime})^{3} \right] d s^{\prime}.
\end{equation}

In order to obtain $\nu_{1}$, we require that:

$$
\langle \phi_{0} \phi_{1} \rangle_{st} = 0.
$$

That is, up to the first order in $q$, the second moment
of $\phi$ is equal to second order of $\phi_{0}$, 

$$
\langle \phi^{2} \rangle_{st} = \langle \phi_{0}^{2} \rangle_{st}
$$

We have chosen that way to establish $\nu_{1}$ because it resembles
somewhat the method of consecutive elimination of perturbing
terms in the near-identity transformation methods in perturbation
theory, please see Chapter 5 of \cite{KC}.

We find without difficulties that

$$
\langle \phi_{0} \phi_{1} \rangle_{st} = \frac{v}{2 \nu_{0}^{2}}
\left[ \nu_{1} + c_{0} (\alpha - \nu_{0}) + 
\frac{1}{2} c_{0} \lambda \frac{v}{\nu_{0}} \right],
$$

and, therefore,

\begin{equation}
\label{row15}
\nu_{1} = c_{0} \left( \nu_{0} - \alpha - \frac{1}{2} c_{0} \lambda \frac{v}{\nu_{0}} \right).
\end{equation}

If we want to risk a seemingly crude approximation and stop the expansion
already after the first order, we need to set

$$
\nu_{0} + \nu_{1} = \alpha,
$$

which implies the following self-consistent equation for $\nu_{0}$:

\begin{equation}
\label{row16}
\nu_{0} = \alpha + \frac{1}{2} \frac{c_{0}}{c_{0} + 1} \lambda \frac{v}{\nu_{0}}.
\end{equation}

Its solution gives:

\begin{equation}
\label{row17}
\nu_{0} = \frac{1}{2} \left( \alpha \pm \sqrt{\alpha^{2} + 2 \frac{c_{0}}{c_{0} + 1} \lambda v}
\right).
\end{equation}

We need to take the ``$+$'' before the square root; otherwise $\nu_{0}$ would be smaller
than zero. 

The question now appears how we can choose $c_{0}$.
We could, in principle, follow the route known from the linear delta expansion method
\cite{FKR}. This method is inherently variational in its nature, containing usually 
an artificial parameter, say, $\mu$,
which, roughly speaking, plays the role analogous to $c_{0}$. 
One imposes there the reasonable condition that one of the physical quantities of interest (let us 
call it $F$) should exhibit ``minimal sensitivity" for the change of $\mu$. That is,
one should have $\partial F/\partial \mu = 0$. 
If we were to follow that route, our quantity of interest, namely $\langle \phi^{2} \rangle$
would have to have a vanishing derivative with respect to $c_{0}$. However, 
this cannot be the case, as 
the above derivative never vanishes. We could either simply 
proceed to the second order and then check the derivatives, or try to establish
another way to find $c_{0}$. A very attractive technique to establish $c_{0}$ is that
proposed by Marinca and coworkers; it consists in minimizing the ``residual error"
in $\phi_{1}$. Let us observe here, however, that equally well one can impose
conditions based on other error-related criteria, similarily as in the statistical
linearization \cite{Kazakov,Caughey,RS,Socha}. While we appreciate valuable insights of Marinca 
and coworkers, in this work we shall use instead the known fact that in the quantum
field theory and statistical physics the so-called strong coupling limit is very often available
in addition to the weak-coupling one. In our language this means that
both limits $\lambda / \alpha \rightarrow 0$ and $\alpha / \lambda \rightarrow 0$
are available. In our simple system this is of course the case, and we have 
in the limit $\alpha / \lambda \rightarrow 0$: 

\begin{equation}
\langle \phi^{2} \rangle_{st} = \frac{2 \sqrt{6} \Gamma(\frac{3}{4})}{\Gamma(\frac{1}{4})}
\sqrt{\frac{v}{\lambda}},
\end{equation}

where $\Gamma(x)$ is the Euler Gamma function. On the other hand, our approximate 
$\langle \phi^{2} \rangle$ gives, for $\alpha = 0$, 
$\sqrt{2 (c_{0} + 1)}/\sqrt{c_{0}} \sqrt{v/l}$. Hence

$$
c_{0} = \frac{\Gamma(\frac{1}{4})^{2}}{12 \Gamma(\frac{3}{4})^{2} - \Gamma(\frac{1}{4})^{2}}
\approx 2.6965826...
$$

A comparison of exact and approximate first-order resutls with the above $c_{0}$
is contained in Fig. 1. The agreement seems quite spectacular if one takes
into account the very low order of approximation. 

\begin{figure}
\begin{center}
\includegraphics[width=4.4cm, height=6cm]{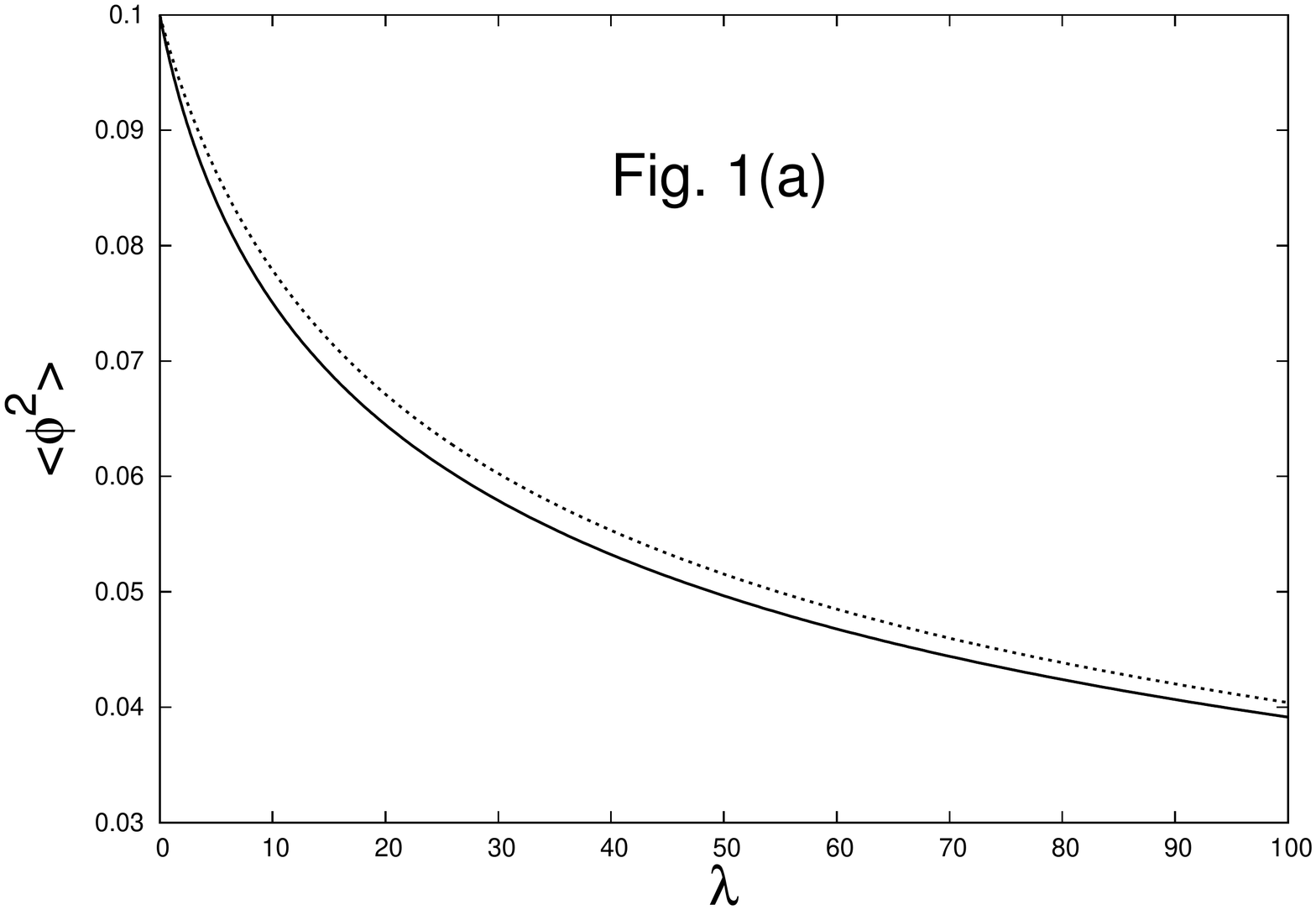}
\includegraphics[width=4.4cm, height=6cm]{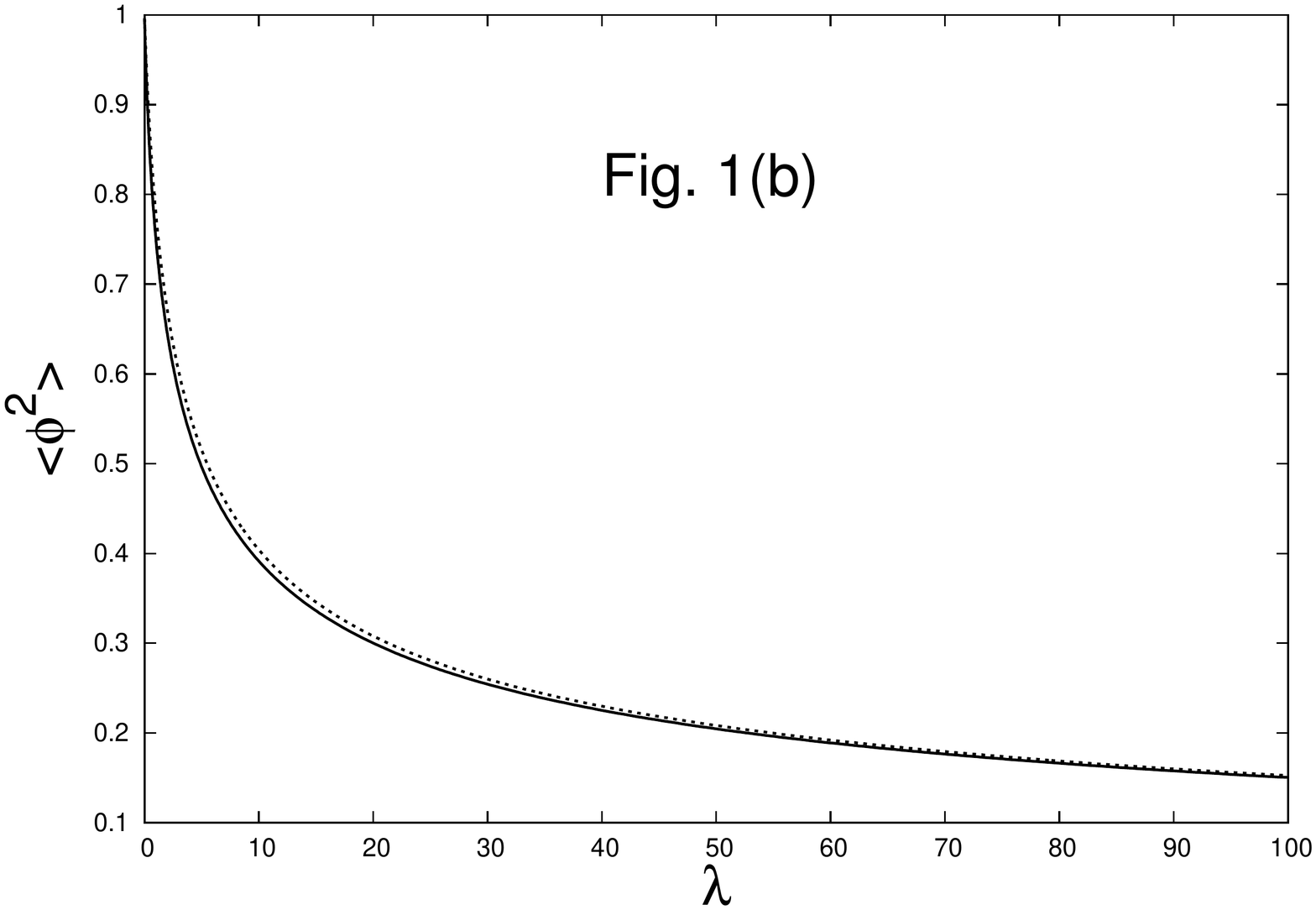}
\includegraphics[width=4.4cm, height=6cm]{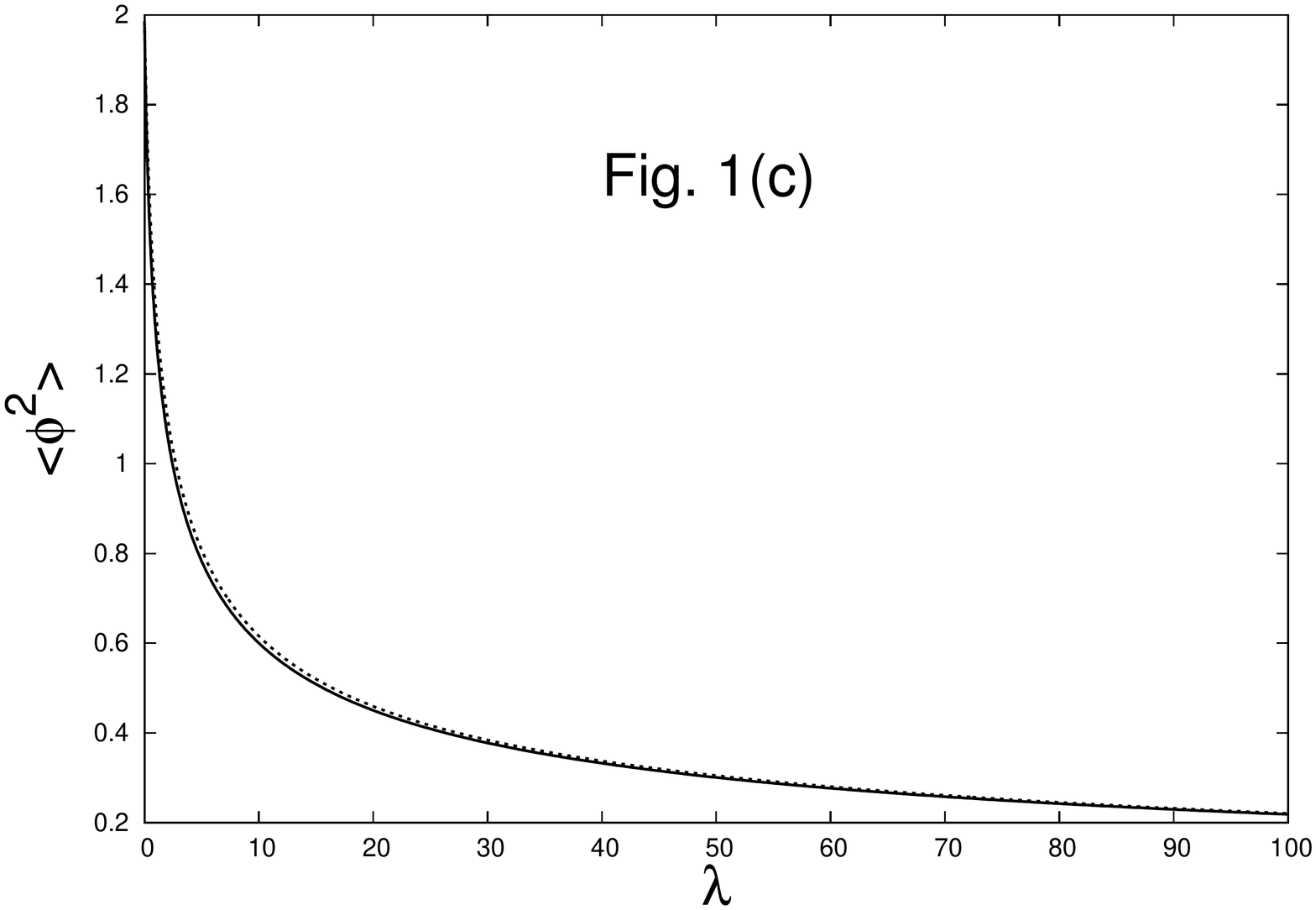}
\caption{Comparison of exact and first-order approximate results for the dependence of 
the second moment of the variable $\phi$ on $\lambda$ for three different values of $v$;
(a) $v = 0.1$; (b) $v = 1.0$; (c) $v = 2.0$. The parameter $\alpha$ is set to $1$.
The solid line represents exact solution, and the dotted line  - the approximate one.}
\end{center}
\end{figure}

Let us observe that for $v$ of the order of $1$ or larger the exact and first-order 
approximate solutions are almost indistinguishable.

Predictably, the results for the fourth moments are less impressive, but still remarkable,
as shown in Fig. 2

\begin{figure}
\begin{center}
\includegraphics[width=4.4cm, height=6cm]{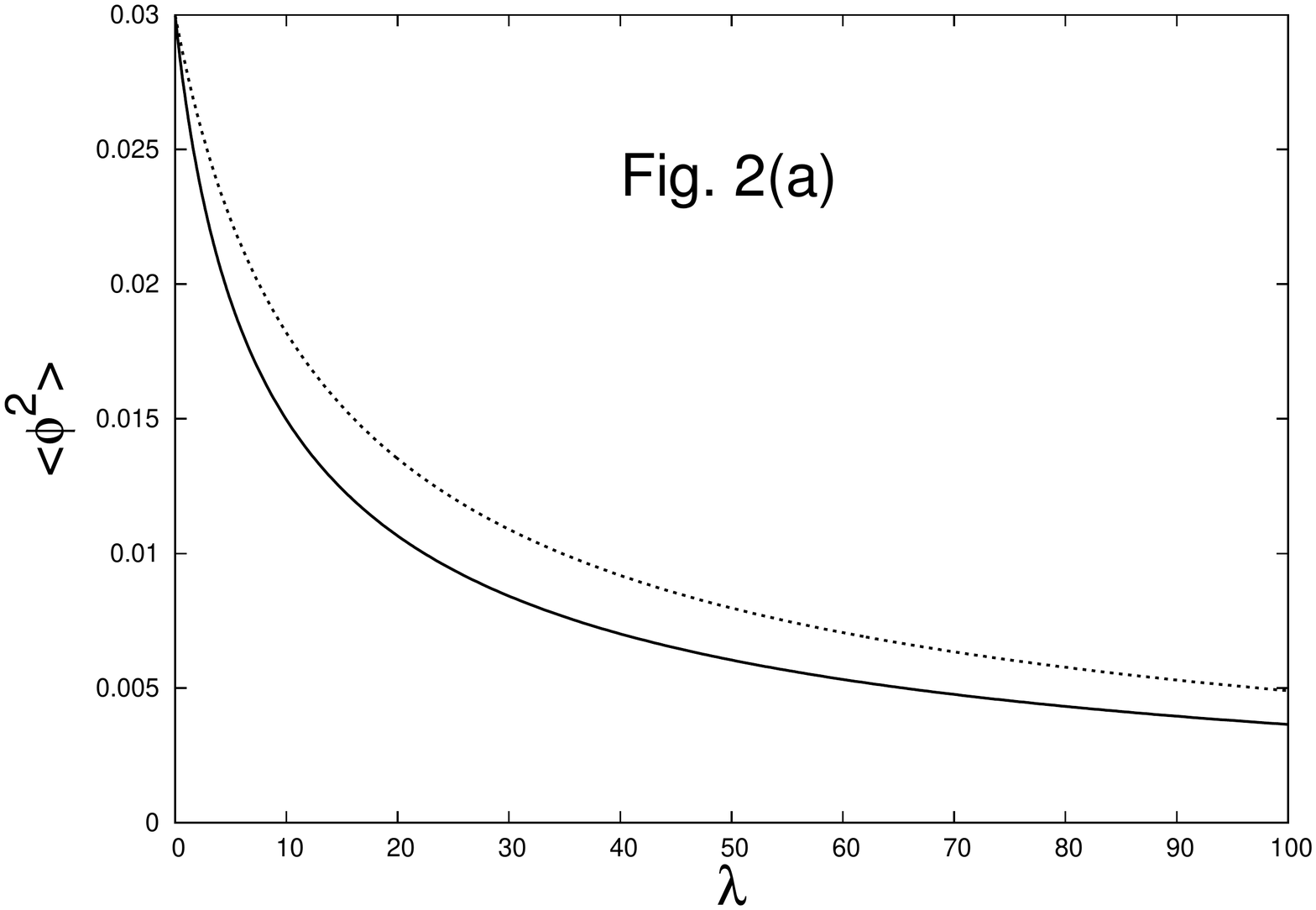}
\includegraphics[width=4.4cm, height=6cm]{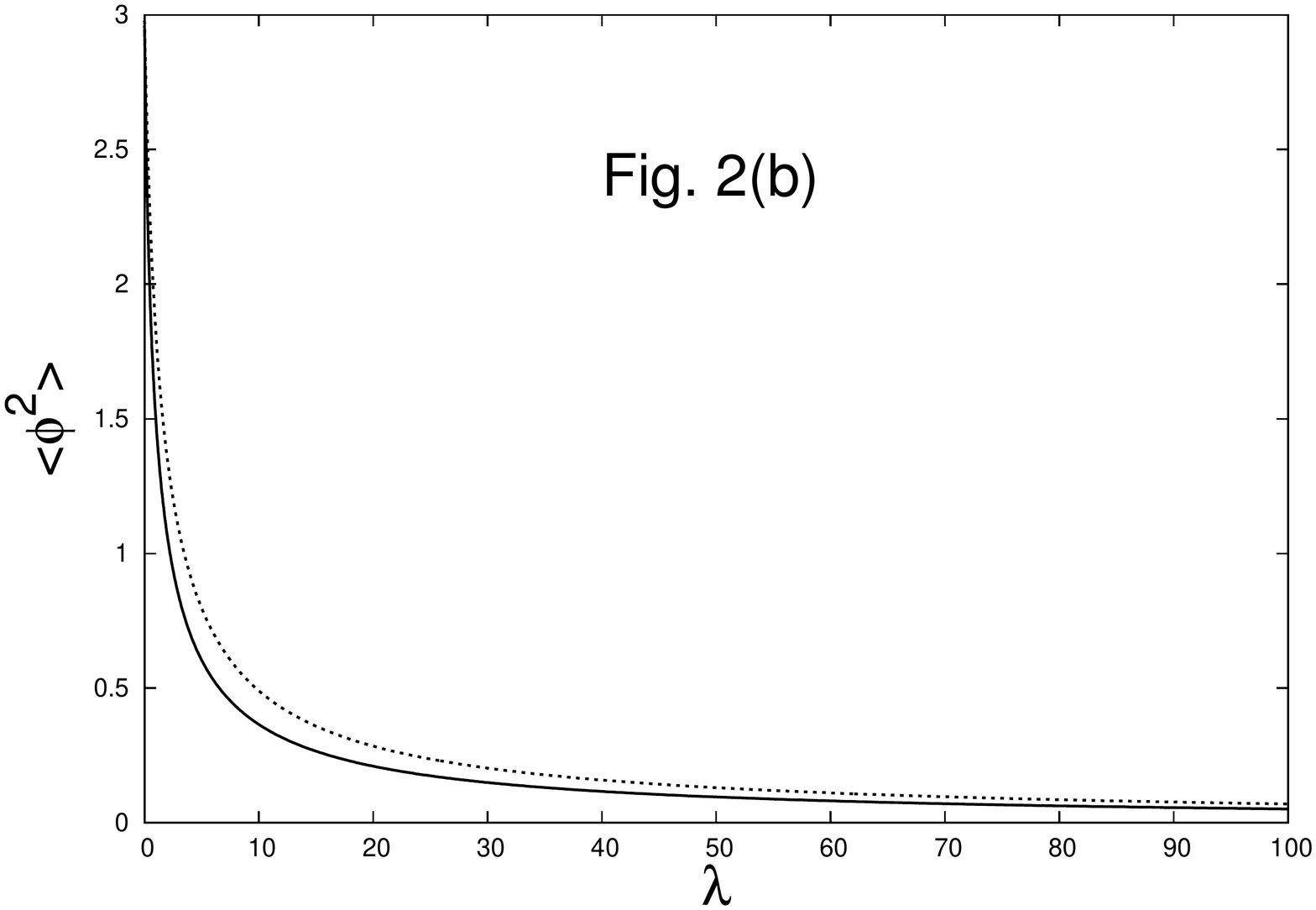}
\includegraphics[width=4.4cm, height=6cm]{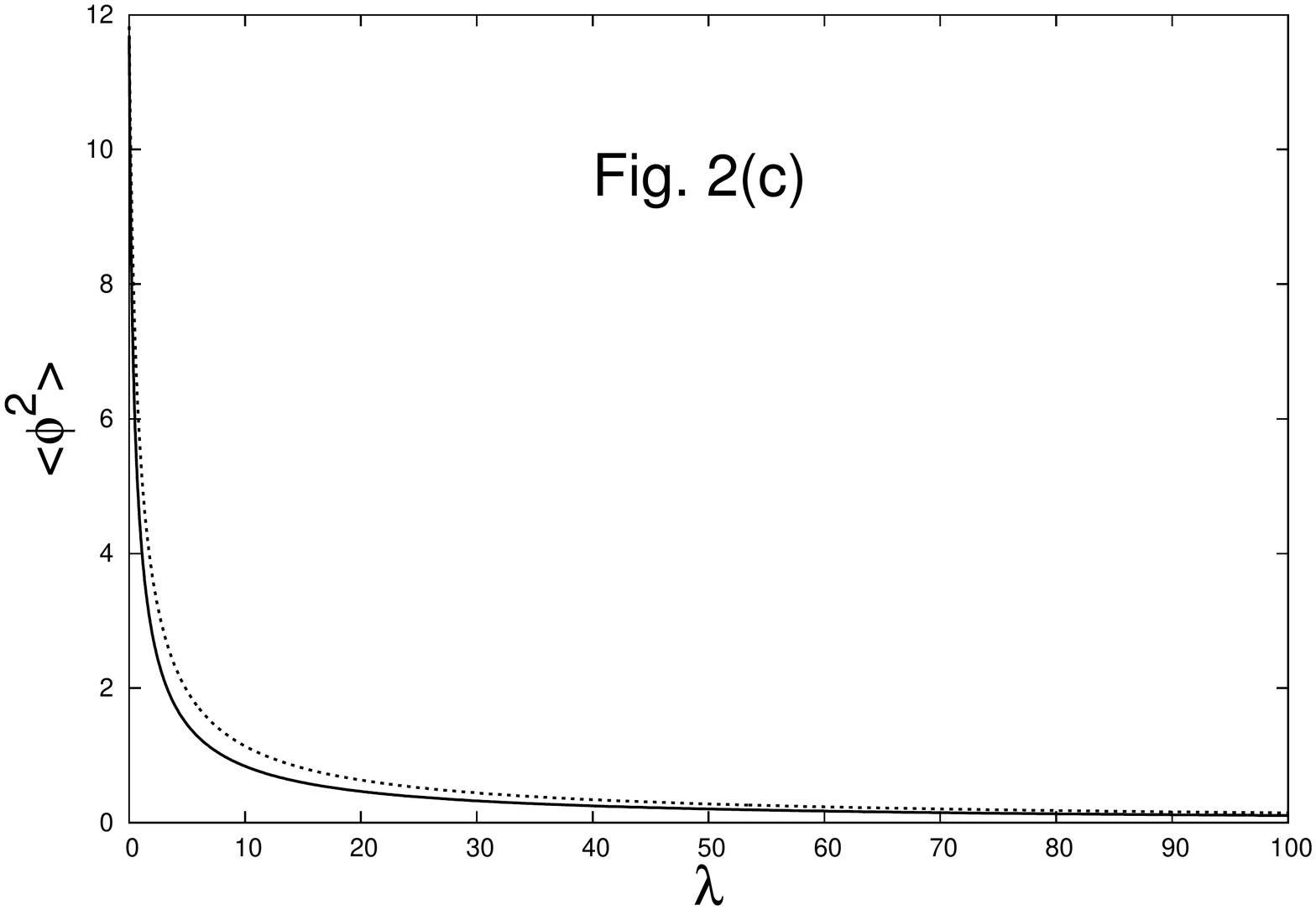}
\caption{Comparison of exact and first-order approximate results for the dependence of 
the fourth moment of the variable $\phi$ on $\lambda$ for three different values of $v$;
(a) $v = 0.1$; (b) $v = 1.0$; (c) $v = 2.0$. The parameter $\alpha$ is set to $1$.
The solid line represents exact solution, and the dotted line - the approximate ones.}
\end{center}
\end{figure}

The results for the fourth moments vary from only qualitatively correct for small values
of $v$ to good agreement with exact values for $v > \alpha$.

Let us proceed to the second order. Now, we need to solve:

\begin{eqnarray}
\label{row18}
\left(\frac{d}{d s} + \nu_{0} \right) \phi_{2}(s) &=& (\nu_{1} + c_{0} (\alpha - \nu_{0})) \phi_{1}(s)
+ \\ \nonumber
&+&
(\nu_{2} - c_{0} \nu_{1} + c_{0} (1 - c_{0}) (\alpha - \nu_{0})) \phi_{0}(s)
+ \\ \nonumber 
&+& 
\frac{1}{6} c_{0} (1 - c_{0}) \lambda \phi_{0}(s)^{3} + 
\frac{1}{2} c_{0} \lambda \phi_{0}(s)^{2} \phi_{1}(s).
\end{eqnarray}

The solution is again quite simple:

\begin{eqnarray}
\label{row19}
\phi_{2}(s) &=& -\int_{0}^{s} e^{-\nu_{0}(s - s^{\prime})} \left[ \left( \nu_{1} + c_{0} (\alpha - \nu_{0})
\right) \phi_{1}(s^{\prime}) + \right. \\ \nonumber 
&+&
\left. \left( \nu_{2} + c_{0} (\alpha - \nu_{0}) - c_{0} (\nu_{1} + c_{0} (\alpha - \nu_{0})) 
\right) \phi_{0}(s^{\prime}) + \right.\\ \nonumber 
&+& 
\left. \frac{1}{6} c_{0} (1 - c_{0}) \lambda \phi_{0}(s^{\prime})^{3} + 
\frac{1}{2} c_{0} \lambda \phi_{0}(s^{\prime})^{2} \phi_{1}(s^{\prime}) \right] d s^{\prime}.
\end{eqnarray}

We require that:

\begin{equation}
\label{row20}
\langle \phi^{2} \rangle_{st} \approx \langle \phi_{0}^{2} \rangle_{st},
\end{equation}

which means that 

$$
2 \langle \phi_{0} \phi_{2} \rangle_{st} + \langle \phi_{1}^{2} \rangle_{st} = 0. 
$$ 

The following algebra is simple but somewhat boring because $\phi_{1}$ and $\phi_{2}$
are no longer Gaussian random variables even though $\phi_{0}$ is. 
It leads to the following self-consistent equations for $\nu_{0}$:

\begin{equation}
\label{row21}
\nu_{0} = \alpha + \frac{c_{0}}{2 c_{0} + 1} \frac{v \lambda}{\nu_{0}} - 
\frac{1}{6} c_{0}^{2} \lambda^{2} \frac{v^{2}}{\nu_{0}^{3}}
\end{equation}

All higher-order moment can, of course, be expressed with the help of $v$ and $\nu_{0}$
because $\phi_{0}$ is Gaussian.

Now, the following difficulty appears. 
If we try to impose the condition: $\nu_{0} + \nu_{1} + \nu_{2} = \alpha$,
which is suitable in the second order, we are confronted with the
algebraic equation of the fourth degree. While we of course can solve it,
it is by no means clear whether the solutions contain a subset of real ones 
for sufficiently broad range of the parameter $c_{0}$. It is actually {\em not}
the case for Eq. (\ref{row21}). In addition, it is not possible to reach
the limit $\lambda/\alpha \rightarrow \infty$ in the sense that there 
are no real solutions for $c_{0}$. Therefore, we have adopted the 
following approach. The function of $\nu_{0}$ which stand on the right-hand
side of (\ref{row21}) can be understood as consisting just first few 
terms of a series in powers of $\nu_{0}^{-1}$. We attempted to sum the
series using the Pade $[1/2]$ approximant.  The result is

\begin{equation}
\nu_{0} = {{3\,{c_0}^2\,l\, \nu_0\,v+\left(6\,\alpha\,{c_0}^2+3\,\alpha\,
 c_0\right)\,{\nu_0}^2+\left(-4\,\alpha^2\,{c_0}^2-4\,\alpha^2\,
 {c_0}-\alpha^2\right)\,{\nu_0}}\over{\left(2\,{c_0}^2+{c_0}
 \right)\,l\,v+\left(6\,{c_0}^2+3\,{c_0}\right)\,{\nu_0}^2+
 \left(-4\,\alpha\,{c_0}^2-4\,\alpha\,{c_0}-\alpha\right)\,{\nu_0}}}.
\end{equation}

This way we obtain for $\nu_{0}$ an algebraic equation of the {\it third} degree.
What is more, one solution is trivial, and it is fairly easy to pick up
the proper one from the remaining two (of course, it is to be positive).
Then the strong-coupling limit is also achievable. Interestingly, we have 
found a {\it negative} value of $c_{0}$, namely, $-1.762159205046485$,
as that for which the limit of large $\lambda$ for the second moment
is obtained. The resulting dependence of the second moments on $\lambda$ 
is shown in Fig. 3.

\begin{figure}
\begin{center}
\includegraphics[width=4.4cm, height=6cm]{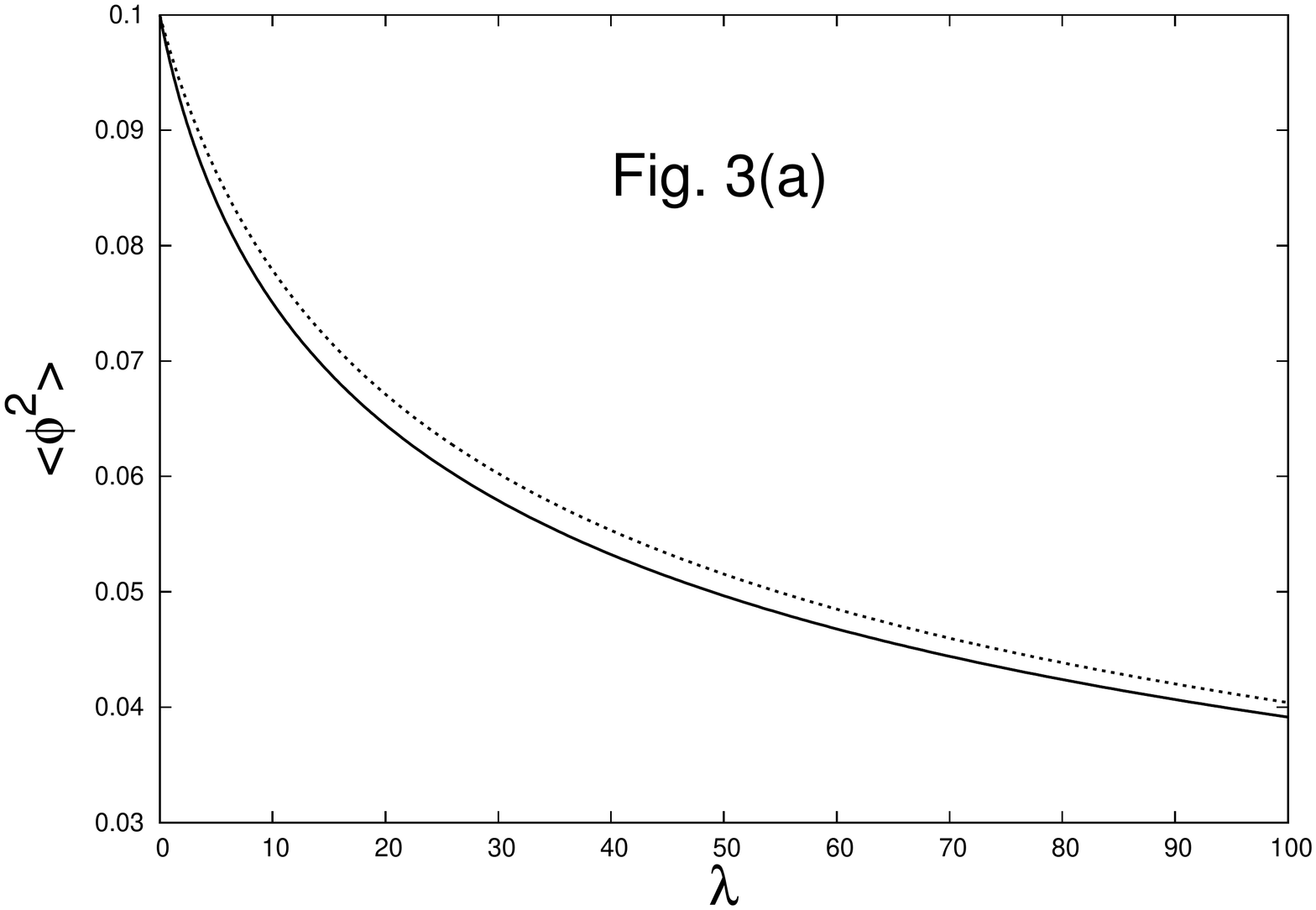}
\includegraphics[width=4.4cm, height=6cm]{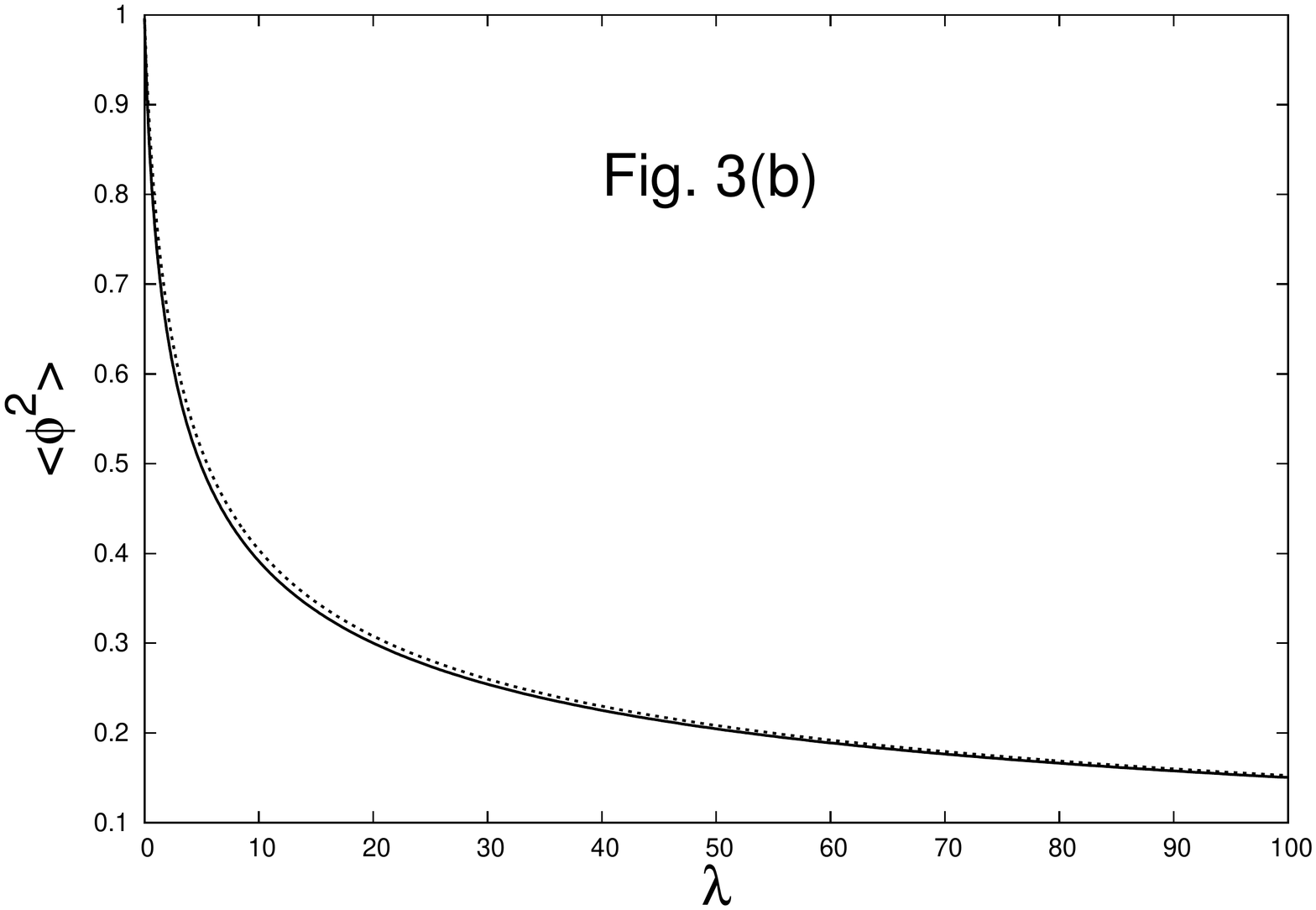}
\includegraphics[width=4.4cm, height=6cm]{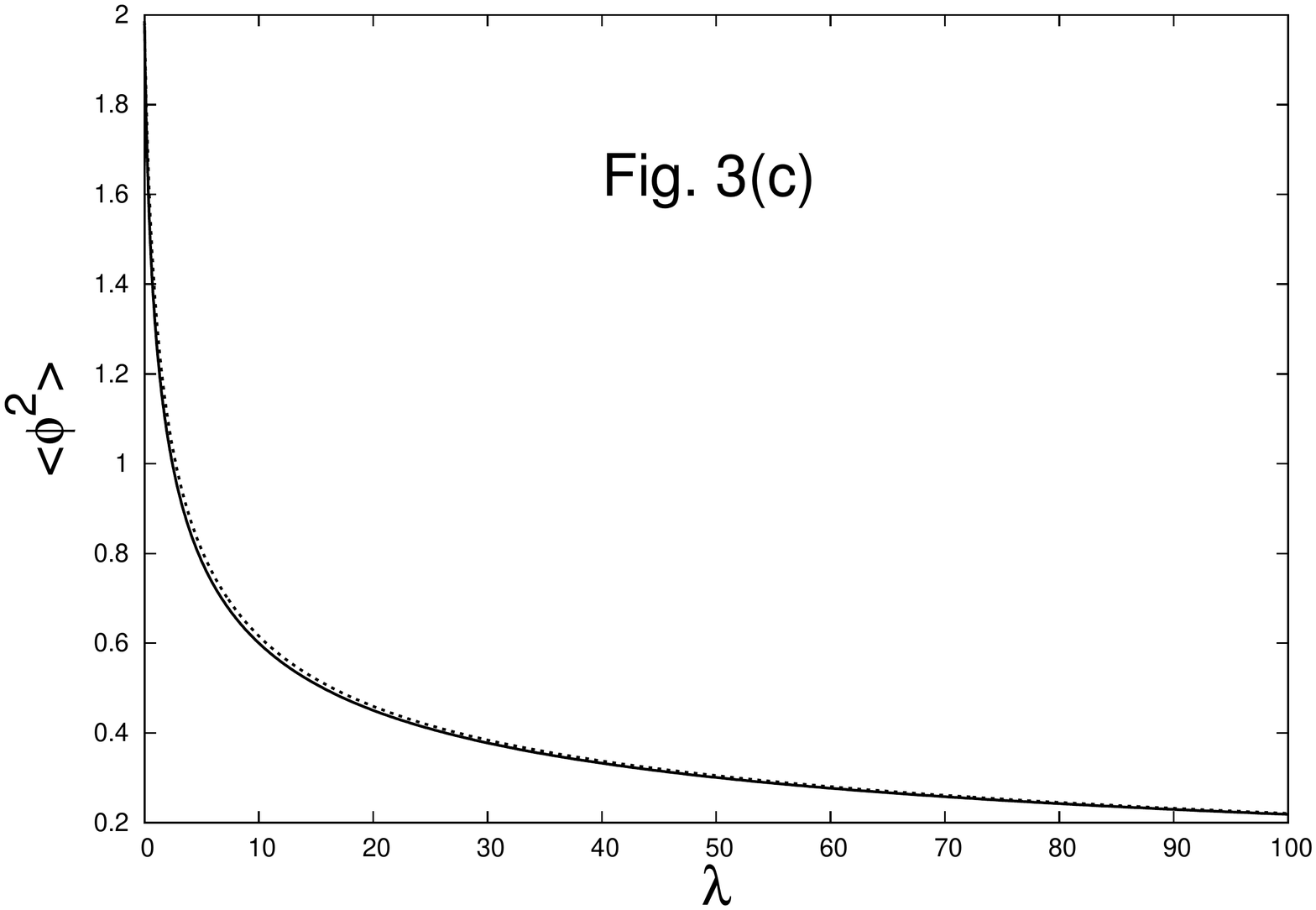}
\caption{Comparison of exact and first-order approximate results for the dependdence of 
the fourth moment of the variable $\phi$ on $\lambda$ for three different values of $v$;
(a) $v = 0.1$; (b) $v = 1.0$; (c) $v = 2.0$. The parameter $\alpha$ is set to $1$.
The solid line represents exact solution, and the dotted line - 
the approximate one.}
\end{center}
\end{figure}

\section{Homotopy analysis method for solution of the Fokker-Planck equation}

To every stochastic equations of the form:

$$
\frac{d \phi}{d s} = -\alpha \phi - \frac{1}{6} \lambda \phi^{3} + f(s)
$$

corresponds the following Fokker-Planck equations of the form:

\begin{equation}
\label{row23}
\frac{\partial \rho}{\partial s} = 
\frac{\partial^{2} \rho}{\partial \phi^{2}} + \frac{1}{v} 
\frac{\partial}{\partial \phi} \left[
\left( \alpha \phi + \frac{1}{6} \lambda \phi^{3} \right) \rho\right]
\end{equation}

where $\rho$ is the distribution function of the random variable $\phi$.
Here, we are interested in the stationary ($s$-independent) solution
of the Fokker-Planck equation. It is, of course, immediately available
and given by (\ref{row5b}). In order to check validity of the homotopy
analysis method we will apply it the Fokker-Planck equations and 
determine the approximate solution.

Alongside Eq. (\ref{row23}) with left-hand side equal to zero we consider the equation:

\begin{eqnarray}
\label{row24}
&&(1-q) \left[ 
\frac{\partial^{2} \rho}{\partial \phi^{2}} + \frac{1}{v}
\frac{\partial \rho}{\partial \phi} \left[
\left( \nu \phi \right) \rho\right] \right] + \\ \nonumber
&+& q c_{0} \left[ 
\frac{\partial^{2} \rho}{\partial \phi^{2}} + \frac{1}{v}
\frac{\partial}{\partial \phi} \left(
\left( \alpha \phi + \frac{1}{6} \lambda \phi^{3} \right) \rho\right)
\right] = 0.
\end{eqnarray}

We look for the solution of (\ref{row24}) in terms of the expansion:

$$
\rho = \rho_{0} + q \rho_{1} + q^{2} \rho_{2} + ...
$$

with analogous expansion of $\nu$, $\nu = \nu_{0} + q \nu_{1} + q^{2} \nu_{2} + ...$.

In the zeroth-order with respect to $q$ we obtain:

\begin{equation}
\label{row25}
\frac{\partial^{2} \rho_{0}}{\partial \phi^{2}} + \frac{1}{v} \frac{\partial}{\phi}
\left( \nu_{0} \phi \rho_{0} \right) = 0.
\end{equation}

The solution which vanishes at $\pm \infty$ is, obviously,

\begin{equation}
\label{row26}
\rho_{0} = N \exp(-\frac{1}{2} \frac{\nu_{0}}{v} \phi^{2}),
\end{equation}

where $N$ is a normalization constant. The differential equation for $\rho_{1}$ is

\begin{equation}
\label{row27}
\frac{d^{2} \rho_{1}}{d \phi^{2}} + 
\frac{1}{v} \frac{d}{d \phi} \left( \nu_{0} \phi \rho_{1}\right)
= -\frac{1}{v} \frac{d}{d \phi} \left(
((c_{0} (\alpha - \nu_{0}) + \nu_{1}) \phi + \frac{1}{6} c_{0} \lambda \phi^{3})
\rho_{0} \right).
\end{equation}

Integrating once and requiring that the integration constant be zero (otherwise
the solution would not be square-integrable) leads to
the simple expression:

\begin{equation}
\label{row28}
\frac{d \rho_{1}}{d \phi} + \frac{1}{v} \left( \nu_{0} \phi \rho_{1} \right)
= \frac{-1}{v} \left[ ((c_{0} (\alpha - \nu_{0}) + \nu_{1}) \phi + 
\frac{1}{6} \lambda c_{0}\phi^{3} ) \rho_{0}
\right] = 0.
\end{equation}

Let us now represent $\rho_{1}$ as the product of $\rho_{0}$ and certain ${\bar \rho}_{1}$,
$\rho_{1} = \rho_{0}{\bar \rho}_{1}$.

Then for ${\bar \rho}_{1}$ we obtain simply:

\begin{equation}
\label{row29}
\frac{d {\bar \rho}_{1}}{d \phi} = -\frac{1}{v} \left[
(c_{0} (\alpha - \nu_{0}) + \nu_{1}) \phi + \frac{1}{6} c_{0} \lambda \phi^{3}
\right].
\end{equation}

Thus,

\begin{equation}
\label{row30}
{\bar \rho}_{1} = \frac{-1}{v} \left[ \frac{1}{2} (c_{0} (\alpha - \nu_{0}) + \nu_{1})
\phi^{2} + \frac{1}{24} c_{0} \lambda \phi^{4} \right].
\end{equation}

If we want to stop the expansion already after the calculcations of the first-order solution,
we need a relation between $\nu_{1}$ and $\nu_{0}$. We impose the condition that, up to the first
order, the second moment of $\phi$ is given solely by $\rho_{0}$, that is:

\begin{equation}
\label{row31}
\int_{-\infty}^{\infty} \phi^{2} \rho_{1}(\phi) = 0.
\end{equation}

This leads to the following relation between $\nu_{1}$ and $\nu_{0}$:

\begin{equation}
\label{row32}
\nu_{1} + c_{0} ( \alpha - \nu_{0} ) = -\frac{5}{12} c_{0} \lambda \frac{v}{\nu_{0}},
\end{equation}

or, taking into account that $\nu_{0} + \nu_{1} = \alpha$, to

\begin{equation}
\label{row33}
\nu_{0} = \frac{1}{2} \left(\alpha \pm \sqrt{\alpha^{2} + \frac{5}{3} \frac{c_{0}}{c_{0}} \lambda v} \right).
\end{equation}

On the other hand, if we require that $\phi$ and $d {\bar \rho_{1}}/d \phi$ are orthogonal in the 
sense of the scalar product

$$
\left( (.), (.) \right) = \int_{-\infty}^{\infty} (.) (.) \rho_{0} d\phi,
$$

then $\nu_{0}$ is given by the same formula as in the previous Section:

$$
\nu_{0} = \frac{1}{2} \left(\alpha \pm \sqrt{\alpha^{2} + 2 \frac{c_{0}}{c_{0}+1} \lambda v} \right).
$$

Remarkably, the two first-order expressions of $\nu_{0}$ are very similar but not identical.

For $\rho_{2}$ we obtain the following equation:

\begin{eqnarray}
\label{row34}
&&\frac{d^{2} \rho_{2}}{d \phi^{2}} + \frac{1}{v} \frac{d}{d \phi} (\nu_{0} \phi \rho_{2})
+ \frac{1}{v} \frac{d}{d \phi} (\nu_{1} \phi \rho_{1}) + 
\frac{1}{v} \frac{d}{d \phi} (\nu_{2} \phi \rho_{0}) + \\ \nonumber
&+&
c_{0} \left[ \frac{d^{2} \rho_{0}}{d \phi^{2}} + \frac{1}{v} \frac{d}{d \phi} 
((\alpha \phi + \frac{\lambda}{6} \phi^{3}) \rho_{0}) \right] + \\ \nonumber
&+&
c_{0} \left[ \frac{d^{2} \rho_{1}}{d \phi^{2}} + \frac{1}{v} \frac{d}{d \phi} 
((\alpha \phi + \frac{\lambda}{6} \phi^{3}) \rho_{1}) \right] = 0
\end{eqnarray}

Simple manipulations and one integration over $\phi$ leads to the following
expression:

\begin{eqnarray}
\label{row35}
\frac{d \rho_{2}}{d \phi} &=& -\frac{1}{v} (\nu_{0} \phi \rho_{2}) 
- \frac{1}{v} [(\nu_{1} + \nu_{0} (\alpha - \nu_{0}) \phi + \frac{1}{6} c_{0}
\lambda \phi^{3}] \rho_{1} - \\ \nonumber
&-&
\frac{1}{v} [(\nu_{2} + c_{0}(\alpha - \nu_{0}) - c_{0} (\nu_{1} + c_{0} (\alpha - \nu_{0})) \phi
\rho_{0}
+ \frac{1}{6} (c_{0} - c_{0}^{2}) \lambda \phi^{3} \rho_{0}].
\end{eqnarray}

On writing $\rho_{2} = \rho_{0} {\bar \rho}_{2}$ we obtain

\begin{eqnarray}
\label{row36}
\frac{d {\bar \rho}_{2}}{d \phi} &=&  
- \frac{1}{v} [(\nu_{1} + \nu_{0} (\alpha - \nu_{0}) \phi + \frac{1}{6} c_{0}
\lambda \phi^{3}] {\bar \rho}_{1} - \\ \nonumber
&-&
\frac{1}{v} [(\nu_{2} + c_{0}(\alpha - \nu_{0}) - c_{0} (\nu_{1} + c_{0} (\alpha - \nu_{0})) \phi
+ \frac{1}{6} (c_{0} - c_{0}^{2}) \lambda \phi^{3}].
\end{eqnarray}

If we now require that $\phi$ is orthogonal to both $d {\bar \rho}_{1}/d \phi$ and
$d{\bar \rho}_{2}/d \phi$ with respect to the scalar product defined above,
we obtain the relation

$$
\nu_{2} + c_{0} (\alpha - \nu_{0}) + \frac{1}{2} c_{0} \lambda \frac{v}{\nu_{0}} = 
\frac{1}{6} c_{0}^{2} \lambda^{2} \frac{v^{2}}{\nu_{0}^3}.
$$ 

What is more, if we decide to break the expansion in terms of $q$ at the second order
so that $\nu_{0} + \nu_{1} + \nu_{2} = \alpha$, we get the implicit (self-consistent)
expression for $\nu_{0}$:

\begin{equation}
\label{row37}
\nu_{0} = \alpha + \frac{c_{0}}{2 c_{0} + 1} \lambda \frac{v}{\nu_{0}}
- \frac{1}{6} \frac{c_{0}^{2}}{2 c_{0} + 1} \frac{v^{2}}{\nu_{0}^{3}} 
\end{equation}

which is identical with that formerly obtained directly from the Langevin equation.

The solution to Eq. (\ref{row36}) can be written as:

\begin{eqnarray}
\label{row38}
{\bar \rho}_{2} &=&
\frac{1}{v^{2}} [\frac{1}{8} (\nu_{1} + c_{0} (\alpha - \nu_{0}))^{2} \phi^{4} +
\frac{1}{48} c_{0} \lambda (\nu_{1} + c_{0} (\alpha - \nu_{0})) \phi^{6} + \\ \nonumber
&+&
\frac{1}{1152} c_{0}^{2} \lambda^{2} \phi^{8}] - 
\frac{1}{v}[\frac{1}{2} (\nu_{2} + c_{0} (\alpha - \nu_{0}) - c_{0}
(\nu_{1} + c_{0} (\alpha - \nu_{0}))) \phi^{2}] + \frac{1}{24} (c_{0} - c_{0}^{2})
\lambda \phi^{4}]
\end{eqnarray}

Let us now require that only $\rho_{0}$ be needed to obtain the second moment of $\phi$,
so that both equations

$$
\int_{-\infty}^{\infty} \phi^{2} \rho_{1} d \phi = 0, 
$$

and 

$$
\int_{-\infty}^{\infty} \phi^{2} \rho_{2} d \phi = 0
$$

hold. In that case we obtain the following relation between $\nu_{2}$ and $\nu_{0}$:

$$
\nu_{2} + c_{0}(\alpha - \nu_{0}) + \frac{5}{12} c_{0} \lambda \frac{v}{\nu_{0}}
= \frac{5}{32} c_{0}^{2} \lambda^{2} \frac{v^{2}}{\nu_{0}^{3}},
$$

and if we want to finish our expansion at the second order, $\nu_{0}$ has to satisfy:

\begin{equation}
\label{row39}
\nu_{0} = \alpha + \frac{5}{6} \frac{c_{0}}{2 c_{0} + 1} \frac{v}{\nu_{0}} -
\frac{5}{32} \frac{c_{0}^{2}}{2 c_{0} + 1} \lambda^{2} \frac{v^{2}}{\nu_{0}^{3}}
\end{equation}

Let us notice that it is only the numerical coefficients which make the 
Eq. (\ref{row39}) slightly different from (\ref{row21}). When we applied
the same procedure, i.e. Pade resummation, to the right-hand side of (\ref{row39}),
solve for $\nu_{0}$ and take the limit of $\lambda \rightarrow \infty$
to obtain $c_{0}$ the resulting dependence of the second moment of $\phi$ 
on $\lambda$ differs only very slightly from that obtained from (\ref{row21}).

\section{Concluding remarks}
In this paper we have applied the homotopy analysis method to a nonlinear stochastic
differential equation with Gaussian-Markovian stochastic force as well as to the 
corresponding Fokker-Planck equation. The value of an artificial parameter the presence 
of which is characteristic for the method has been fixed by taking the limit of strong
nonlinearity. Approximate solutions have been obtained
and compared with the exact ones. It has been demonstrated that the second moments
of the dependent stochastic variables as given by the homotopy analysis method
agree very remarkably with exact moments. Broad perspectives of applications
for the method seem to be open; for, despite the simplicity of the considered model,
it possesses some characteristics of the stochastic differential equations 
of statistical mechanics and quantum field theory in the so-called stochastic
quantization.

Work is in progress on application of both those techniques in the physics of cold atomic 
gases,
and quantum optics.

\bibliographystyle{els    psi_int_star = np.conjugate(psi_int)article-num}
\bibliography{<your-bib-database>}



\end{document}